\documentclass{PoS}

\usepackage[sort&compress,numbers]{natbib}
\usepackage{mathtools}
\usepackage{amsfonts} 
\usepackage{amssymb} 
\usepackage{amsmath} 
\usepackage{graphicx} 
\usepackage{latexsym} 
\usepackage{verbatim} 

\newcommand{\K}[0]{\mathcal K}
\newcommand{\Kdfth}[0]{\mathcal K_{\mathrm{df},3}}
\newcommand{\PV}[0]{\widetilde{\mathrm{PV}}}

\title{Three-particle quantization condition: an update}

\ShortTitle{Three-particle quantization condition: an update}

\author{Maxwell T. Hansen\thanks{Present address: Institut f\"ur Kernphysik and
Helmholz Institute Mainz, Johannes Gutenberg-Universit\"at Mainz, 
55099 Mainz, Germany}\\
        Physics Department, University of Washington, Seattle, WA 98195-1560, USA\\
        E-mail: \email{mth28@uw.edu}}

\author{\speaker{Stephen R. Sharpe}\\
        Physics Department, University of Washington, Seattle, WA 98195-1560, USA\\
        E-mail: \email{srsharpe@uw.edu}}

\abstract{We give an update on our derivation of a quantization
condition relating the finite-volume spectrum of three particles
in a cubic box to infinite-volume scattering quantities.
We have discovered and
fixed technical problems in the derivation sketched in 
the proceedings of last year's lattice conference~[1],
and have presented a detailed description of the corrected
derivation in Ref.~[2].
Here we give an overview of the problems and their solutions, and
describe open questions.}

\FullConference{The 32nd International Symposium on Lattice Field Theory,\\
		23-28 June, 2014\\
		Columbia University New York, NY}

\begin{document}

\section{Introduction}
In this talk we give an update on our derivation of a
quantization condition relating the finite-volume (FV) spectrum
in the three-particle sector to infinite-volume scattering
quantities. At last year's lattice conference we presented a
preliminary result~\cite{Hansen:2013dla}. This, however, turned out not
to be fully correct, due to technical issues in the derivation.
We have recently finished a corrected derivation,
written up at length in Ref.~\cite{Hansen:2014eka}. 
We aim to give here an overview of the
result and a sketch of the technical issues that have arisen.

Many resonances have significant coupling to three-particle channels,
e.g. $\omega\to3\pi$, $K^*\to K\pi\pi$ and $N(1440)\to N\pi\pi$,
and if we wish to determine the properties of such resonances from
first principles using lattice QCD, a formalism connecting the
FV spectrum to infinite-volume (measurable) quantities
is needed.\footnote{%
Indeed, the recent coupled-channel analysis of Ref.~\cite{Dudek:2014qha} 
is limited at the 
highest energies by the lack of a three-particle formalism,
a problem that will become more acute as quark masses are lowered
towards their physical values.}
Additionally, three-body interactions are important in
nuclear physics and perhaps also in pion- and kaon-condensed matter.
A formalism is needed to extract these interactions from the FV
spectrum. Finally, interesting weak decays involve three particles,
e.g. $K\to3\pi$, and in order to study these one needs as a first step
to understand the FV effects on the three-particle system.

The three-particle quantization condition is at the frontier 
of finite-volume formalism.
The two-particle quantization condition is well understood. 
This formalism was first developed by 
L\"uscher~\cite{Luscher:1986n2,Luscher:1991n1}, 
and is now widely implemented in numerical calculations. 
By contrast, although important progress has been 
made~\cite{Polejaeva:2012ut,Briceno:2012rv}, 
our current understanding of the three-particle sector is incomplete.

The work reviewed here is applicable to any relativistic field theory
with a single scalar field (of mass $m$), with the only simplification
being that we impose a $Z_2$ symmetry (akin to G-parity) 
such that all vertices have an even number of legs. 
This means that the relevant physical scattering amplitudes are those
for $2\to2$ and $3\to3$ processes, while $2\to 3$ transitions
do not occur. 
We work to all orders in perturbation theory,
and do not assume that couplings are weak.
On the FV side we assume the simplest possible implementation: 
a periodic cubic box with side length $L$.
We expect that extensions to non-identical particles, to spin,
and to asymmetric boxes
will be relatively straightforward using methods developed for
two particles.

\section{Quantization conditions}

It is useful to recall the two-particle single-channel quantization condition
in a form similar to that derived in Ref.~\cite{Kim:2005gf}. 
For fixed total momentum $\vec P$, 
FV states have energies, $E$, such that\footnote{%
This result holds up to exponentially suppressed FV effects
($\propto e^{-mL}$) which we ignore throughout.}
\begin{equation}
\det(F_2^{-1}+i{\cal K}_2)= 0\,.
\label{eq:QC2}
\end{equation}
Here ${\cal K}_2$ is the two-particle K-matrix,
an infinite-volume quantity,
while $F_2$ is a known kinematical factor, dependent on the volume.
Both quantities are matrices in the space of angular momentum
in the two-particle center-of-mass (CM) frame,
e.g. $(F_2)_{\ell_1,m_1;\ell_2,m_2}$.
This space arises because, for given $(E,\vec P)$,
the only degree of freedom left for two on-shell particles
is the direction of momentum for one of the particles in the CM frame.
By decomposing in spherical harmonics, 
this degree of freedom may be equivalently expressed 
as the CM angular momentum. 
While ${\cal K}_2$ is diagonal in angular momentum,
$F_2$ knows about the box shape and mixes angular momenta.
Roughly speaking, $F_2$ is the difference between the FV
momentum sum and infinite-volume integral over a two-particle cut.
It is expressible in terms of L\"uscher zeta functions~\cite{Luscher:1986n2}.
The beauty of Eq.~(\ref{eq:QC2}) is that it separates 
finite and infinite-volume quantities.\footnote{%
The result of Ref.~\cite{Kim:2005gf} is  actually
$\det(F_{i\epsilon}^{-1} + i{\cal M}) = 0$, with ${\cal M}$ the scattering
amplitude and $F_{i\epsilon}$ a modified kinematical factor. 
This form is equivalent to Eq.~(\ref{eq:QC2}), 
as shown in Ref.~\cite{Hansen:2014eka}, 
and the latter is more useful for our purposes.}

The extension to three particles requires additional kinematical variables.
The choice we find convenient divides the particles (arbitrarily) into
a ``spectator'' and the ``scattering pair''.
The variables are then the FV momentum
of the spectator, $\vec k=2\pi \vec n/L$, 
and the angular momentum of the scattering pair in their CM frame.
Thus each matrix ``index'' is written $\vec k,\ell,m$.
With this change, our result for the three-particle
quantization condition can be written in a form that is
superficially similar to the two-particle result (\ref{eq:QC2}), namely
\begin{equation}
\det(F_3^{-1} + i \Kdfth) = 0\,.
\label{eq:QC3}
\end{equation}
However, $F_3$ is not a purely kinematical function as it depends on 
${\cal K}_2$:
\begin{equation}
F_3 = \frac1{2\omega L^3}
\left[-\frac{2F}3 
+ \frac1{F^{-1} + (1+\K G)^{-1} \K F}\right]\,.
\label{eq:QC3a}
\end{equation}
Here all quantities are matrices in the enlarged space.
$\K$ and $F$ are simply ${\cal K}_2$ and $F_2$, respectively,
multiplied by $\delta_{\vec k,\vec k'}$,
the prefactor 
$\omega L^3$ is a diagonal matrix with entries $\sqrt{\vec k^2+m^2}\,L^3$,
and $G$ is a regulated free non-relativistic propagator 
(defined precisely in Ref.~\cite{Hansen:2014eka}).
In principle, $\K$ can be determined by interpolating results obtained
from the two-particle quantization condition, in which case $F_3$
is known and one can use Eq.~(\ref{eq:QC3}) to determine $\Kdfth$.

The forms of Eqs.~(\ref{eq:QC3}) and (\ref{eq:QC3a}) are
(aside from minor changes in notation) 
identical to those given in last year's proceedings 
[Eqs.~(6.1) and (6.2) of Ref.~\cite{Hansen:2013dla}].
There have, however, been two very important changes in the meaning of the symbols.
The first concerns $F$, which is a sum-integral difference. The integral
part is defined using a modified principal value prescription (denoted $\PV$) 
which was incompletely specified in Ref.~\cite{Hansen:2013dla},
but is completely defined in Ref.~\cite{Hansen:2014eka}.

The second change concerns $\Kdfth$ and is more significant.
The good news is that this an infinite-volume three-to-three
scattering quantity, so that the FV spectrum is determined
using Eq.~(\ref{eq:QC3}) in terms of infinite-volume quantities.
The bad news is that, contrary to our claim last year, the relation
of $\Kdfth$ to the three-particle scattering amplitude is not yet clear.
We strongly suspect, however, that such a relation exists, for it does in the
non-relativistic limit~\cite{Polejaeva:2012ut}.
$\Kdfth$ is defined in Ref.~\cite{Hansen:2014eka} as a sum of the same
diagrams which build up the divergence-free three-to-three scattering
amplitude,\footnote{%
``Divergence-free'' indicates that divergence in the physical 
three-to-three scattering amplitude
(which occurs for on-shell above-threshold momenta) is subtracted.
For details see Refs.~\cite{Hansen:2013dla,Hansen:2014eka}. 
This aspect of the problem was understood last year, 
and we do not discuss it further.}
except that the $\PV$ prescription is used, which requires specification
of the ordering of integrations, and that
additional ``decorations'' must be added to the diagrams.

In the following two sections we give some further details concerning
these two changes.

\section{Cusps and their removal with the $\PV$ prescription}

\begin{figure}
\begin{center}
\includegraphics[scale=0.8]{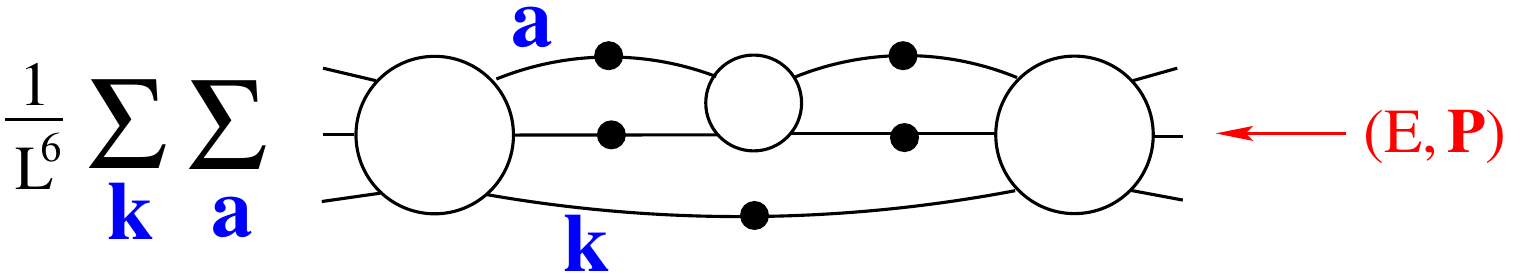}
\caption{Segment of a Feynman diagram contributing to FV correlator. 
Open circles and lines are BS kernels and
fully-dressed propagators, respectively. See text for further discussion.}
\label{fig:cut}
\end{center}
\end{figure}

Our analysis considers a FV correlation function in the
sector with an odd number of particles, evaluated in Minkowski space
as a function of $E$ (and for fixed $\vec P$ and $L$).
Poles in $E$ then give the FV spectrum.
Power-law FV effects in the correlator arise from the differences
between loop sums and integrals over singular (but integrable)
summands. Such singularities occur only when there are
three-particle intermediate states.
An example is the ``cut'' which runs through the momenta $\vec a$ and
$\vec k$ in the diagram of Fig.~\ref{fig:cut}.
Our analysis consists of replacing the FV momentum
sums with infinite-volume integrals plus a residue that brings in the
factors of $F$.
The need for 
the $\PV$ prescription arises from cuts where one or both sides
contains a two-particle Bethe-Salpeter (BS) kernel (so that there is a
spectator). This is true for either cut in Fig.~\ref{fig:cut},
and we focus here on the cut on the left side.

Sums can be replaced by integrals (up to exponentially suppressed
corrections) for non-singular summands. Thus we need consider only 
the singular part of the summand, which can be written
\begin{equation}
\frac1{L^6}\sum_{\vec k} \sum_{\vec a}
\frac{A(\vec k, \vec a) B(\vec k,\vec a)}
{E-\omega_k-\omega_a-\omega_{ka}}
\,.
\end{equation}
Here the sums run over FV momenta,
$A$ and $B$ are smooth functions representing the kernels respectively
to the left and right of the cut, and $\omega_{k}$, $\omega_a$
and $\omega_{ka}$ are the on-shell energies of the three cut particles
[e.g. $\omega_{ka}=\sqrt{(\vec P-\vec k-\vec a)^2+m^2}$].

The first step is to make the replacement
\begin{equation}
L^{-3} \sum_{\vec a} = (2\pi)^{-3} \int d^3\vec a +
\left[L^{-3} \sum_{\vec a} - (2\pi)^{-3} \int d^3\vec a\right]
\,.
\label{eq:split}
\end{equation}
The sum-integral difference leads to the kinematical factor $F$
multiplied by the {\em on-shell} values of $A$ and $B$~\cite{Kim:2005gf}.
To define $F$ we need to specify a pole prescription for the integral,
and, for reasons to be seen shortly, we use $\PV$ instead of the
$i\epsilon$ prescription used in Ref.~\cite{Kim:2005gf}.
For the $F$ term the remaining sum over $\vec k$ must remain 
a sum, since $F$ has singularities as a function of $\vec k$.
This sum over $\vec k$, together with the dependence of $F$ on
the two-particle angular momentum $\ell,m$, builds up the matrix indices
introduced above.

The second step is to deal with the $\vec a$ integral in (\ref{eq:split}).
This is acted on by a sum over $\vec k$, and we would like to replace
this outer sum with an integral. While the integral over $\vec a$ does lead
(for $i\epsilon$, PV and $\PV$ pole prescriptions) 
to a {\em finite} function of $\vec k$,
it does not in general yield a {\em smooth} function.
In particular, there can be cusps when $\vec k$ is such
that the scattering pair is at threshold.\footnote{%
Note that when $|\vec k|$ is large enough, the scattering pair lies
below threshold.}
If cusps are present then the difference between the sum and integral
over $\vec k$ gives rise to power-law dependence on $L$ (beginning at
$1/L^4$) that one must explicitly include to derive the quantization condition.
Our choice is to use a pole prescription which is cusp free,
namely the $\PV$ prescription. Then we can simply replace the sum over
$\vec k$ with an integral.
We stress that this issue arises first for three particles because only
then do we need to consider the analytic properties of the sum-integral
difference below the two-particle threshold.

\begin{figure}
\begin{center}
\includegraphics[scale=0.47]{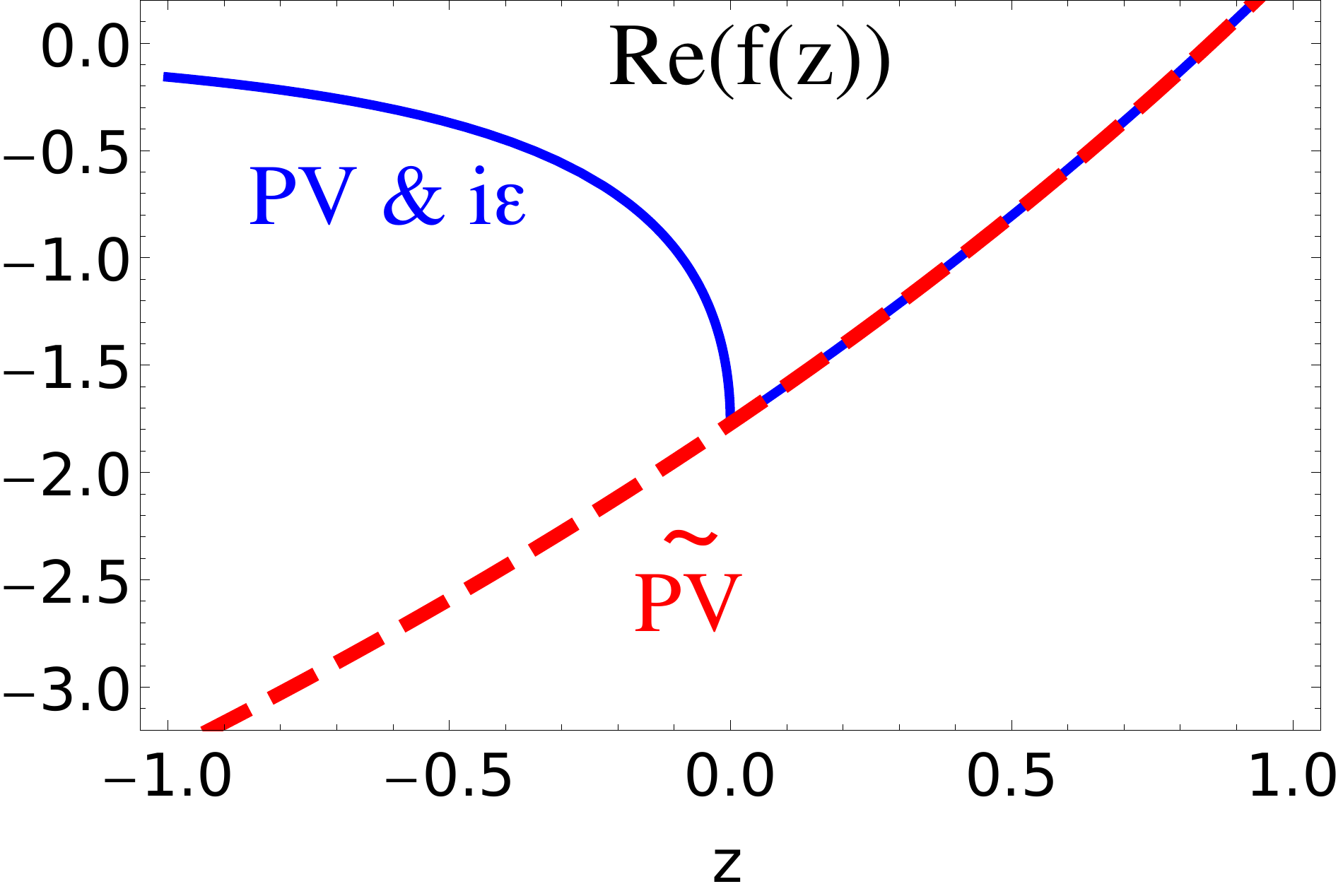}
\hspace{0.2truein}
\includegraphics[scale=0.4]{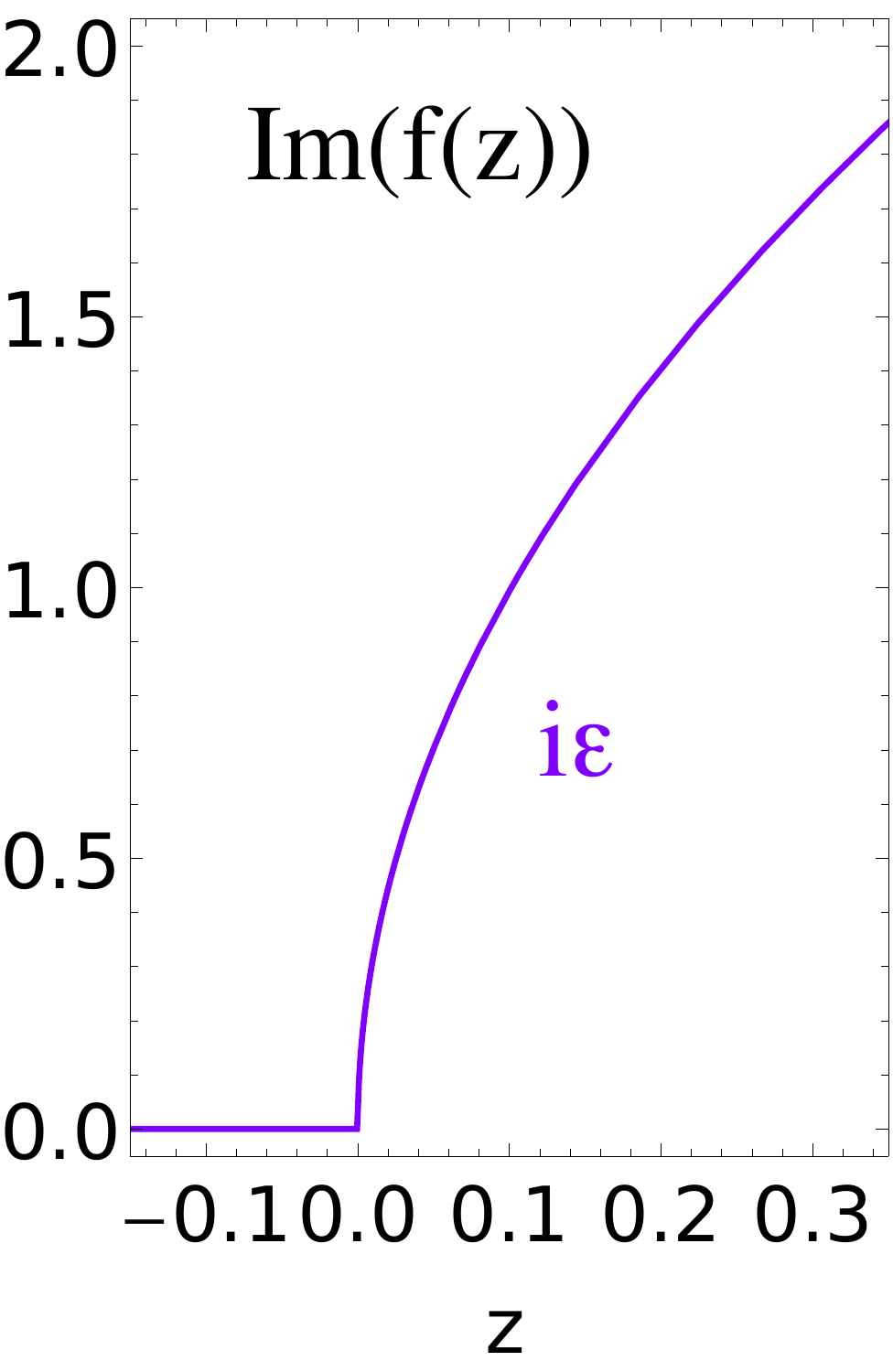}
\caption{Results for the integral of Eq.~(\protect\ref{eq:f[z]}).}
\label{fig:f[z]}
\end{center}
\end{figure}

The presence of cusps is well known using the standard $i\epsilon$ 
prescription, and is indeed required by unitarity.
To illustrate the cusps that we encounter in the present context
and to better explain the $\PV$ prescription
we consider a simple integral that (as explained in App.~B of
Ref.~\cite{Hansen:2014eka}) captures the essential features of
the actual integral:
\begin{equation}
f(z) = \int_0^\infty dx \sqrt{x}\, \frac{e^{z-x}}{z-x}\,.
\label{eq:f[z]}
\end{equation}
Results for $f(z)$ using the $i\epsilon$, standard PV, and
$\PV$ prescriptions are shown in Fig.~\ref{fig:f[z]}.
Observe that both $i\epsilon$ and PV prescriptions lead to
cusps at $z=0$ (which corresponds to the two-particle threshold
in this example). For the $i\epsilon$ prescription cusps appear
in both real and imaginary parts. For the PV prescription,
which is equivalent to the real part of the $i\epsilon$ prescription,
we have only the real cusp.
As we show in Ref.~\cite{Hansen:2014eka}, and is likely well known
in the three-body community (see, e.g., Ref.~\cite{Polejaeva:2012ut}),
the difference between the cusped PV result and the
smooth extension of the $z>0$ curve
(i.e. the red dashed curve in Fig.~\ref{fig:f[z]})
can be calculated analytically.
For the integral of actual interest, the difference depends
on the analytic
continuation of the product of on-shell amplitudes multiplied
by an analytically continued phase-space factor.
The $\PV$ prescription consists in subtracting this difference,
and then smoothly turning off the subtraction far below threshold.
Slightly below threshold, this yields the (unique)
analytic continuation of the above-threshold standard PV result.

Using the $\PV$ prescription throughout, we are able to separate
FV and infinite-volume contributions. In the two-particle subsector
(e.g. when iterating insertions of the two-to-two BS kernel
in Fig.~\ref{fig:cut}) this leads to the appearance of
a two-particle K-matrix, ${\cal K}_2$, rather than the
corresponding scattering amplitude. ${\cal K}_2$ is the standard
K-matrix above threshold (where PV and $\PV$ are equivalent),
but just below threshold corresponds to the analytic continuation of
the above-threshold function. 
For a given value of $\ell$, it is simply
$16 \pi E^*/(a^* \cot\delta_\ell)$ above threshold
(with $E^*$ and $a^*$ respectively
the two-particle CM energy and momentum), 
while the result below threshold is defined 
via the standard threshold expansion.
This is, in fact, the standard choice in the literature
when using the two-particle quantization condition below threshold
to study FV effects on bound states. In this sense, our prescription
is not new (aside from the turn-off of the subtraction far below
threshold).

\section{Enforcing particle-interchange symmetry}

While the $\PV$ prescription avoids the issue of cusps, it 
introduces substantial complications into the definition of $\Kdfth$. 
This is very difficult to explain without going into technicalities,
so here we give only a crude sketch of the issues.
For the full story see Ref.~\cite{Hansen:2014eka}.

\begin{figure}
\begin{center}
\includegraphics[scale=0.8]{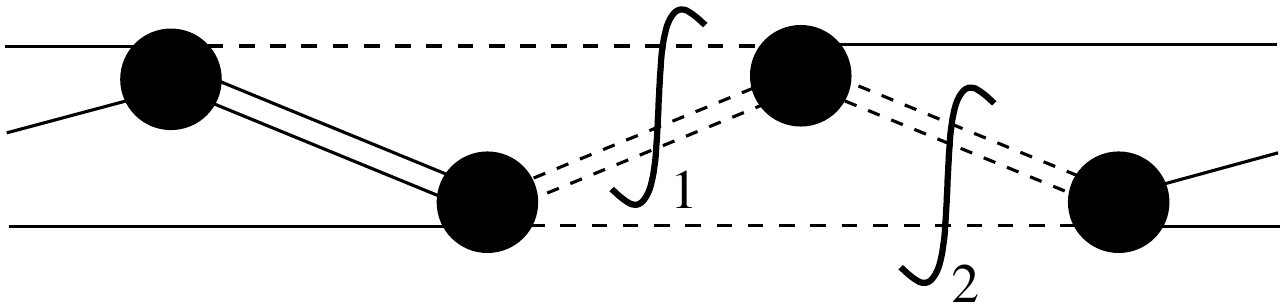}
\caption{Example of a two-loop contribution to $\Kdfth$. Notation as follows:
filled circles: ${\cal K}_2$; 
double dashed lines: divergent part of propagator; 
double solid lines: finite part of propagator; 
single dashed lines: on-shell particle;
single solid lines: amputated on-shell external propagator.
See Ref.~\cite{Hansen:2014eka} for more details.}
\label{fig:K4}
\end{center}
\end{figure}

The basic issue is that the $\PV$ prescription breaks
particle-interchange symmetry. One of the particles
(the spectator) is treated differently from the other two.
One consequence is that in diagrams containing adjacent two-to-two
BS kernels on different pairs (i.e. where the spectator changes)
the order of $\PV$-regulated integrals matters. This is not the case
with the $i\epsilon$ prescription. An example is shown in Fig.~\ref{fig:K4},
which is one of many two-loop contributions to $\Kdfth$.
Here we have performed a summation of two-to-two BS kernels 
into K-matrices.
Our construction requires that the two $\PV$-regulated integrals be
done in the order shown. One also sees here another complication:
defining a divergence-free three-particle scattering quantity requires
that the central propagators are broken up into singular and finite parts.
Our construction gives a prescription for
this breakup, and the resulting order of integrals, for diagrams
of all orders.

A second consequence of the breaking of particle-interchange symmetry
by the $\PV$ prescription is that the resulting infinite-volume quantity,
$\Kdfth$, is not manifestly symmetric. More precisely, while it
is symmetric under interchange of external legs, 
it contains internally some parts that are asymmetric.
We refer the reader to Ref.~\cite{Hansen:2014eka}
for complete details.
Again, a precise formula for these asymmetric parts can be given
for all diagrams. 

Because of the complicated definition of $\Kdfth$, we have,
so far, been unable to determine its relation to the physical
three-to-three scattering amplitude. We have evidence that
such a relation exists close to threshold from the work of
Ref.~\cite{Polejaeva:2012ut}, and thus we are presently
concentrating on the threshold expansion to address this question.

\section{Outlook}

Many issues have not been discussed in this short presentation.
For example, 
how can the results (\ref{eq:QC3}) and (\ref{eq:QC3a}) be checked?
A major check is the agreement of our results near the three-particle
threshold to those obtained from non-relativistic quantum mechanics. 
This was mentioned last year~\cite{Hansen:2013dla} and
we are presently writing up a detailed discussion.
Are these complicated formulae useful in practice?
This we have begun to address in Ref.~\cite{Hansen:2014eka}.
As in the two-particle case, practical utility requires
truncation of the infinite-dimensional matrices to a finite subspace.
We have shown in Ref.~\cite{Hansen:2014eka}
that this can be done in principle,
and have worked out in detail the simplest truncation.
More work is needed here, for example using model amplitudes
to illustrate in detail how the result may be applied.

In closing, we stress again that
the most important unresolved issue is to determine the
relation of $\Kdfth$ to physical scattering amplitudes. 

\section{Acknowledgments}
We thank Ra\'ul Brice\~no, Zohreh Davoudi and Akaki Rusetsky for
helpful discussions. 
This work was supported in part by the United States Department of Energy 
grants DE-FG02-96ER40956 and DE-SC0011637.
MTH was supported in part by the Fermilab Fellowship in Theoretical Physics.
Fermilab is operated by Fermi Research Alliance, LLC, under Contract
No.~DE-AC02-07CH11359 with the United States Department of Energy.


\bibliographystyle{apsrev4-1}
\bibliography{ref}

\end{document}